\documentclass[aps,prd,superscriptaddress,amsmath,amsfonts,groupedaddress,10pt,english]{revtex4}
\usepackage{amssymb,amsmath,epsfig}
\usepackage{graphicx}
\usepackage{float}
\usepackage[caption=false]{subfig}
\usepackage{natbib}
\usepackage{hyperref} 
\usepackage{xcolor}
\usepackage{pifont}

\begin{document}

\title{Parameter Space, Realistic Matter, and Universal Relations in Bose--Einstein Condensate Dark Stars}

\author{M. Ilyas}
\email{ilyas\_mia@yahoo.com}
\affiliation
{Institute of Physics, Gomal University, Dera Ismail Khan, 29220, KP, Pakistan}
\date{}

\begin{abstract}
We study slowly rotating Bose--Einstein condensate (BEC) dark stars by solving the Tolman--Oppenheimer--Volkoff, Hartle dipole, and Postnikov--Hinderer equations together for a polytropic equation of state with a Lee--Huang--Yang correction of strength $\zeta$~\citep{Panotopoulos2026}. A continuous scan of $\zeta$ from 0 to 1.5 shows the mean-field-to-corrected transition is smooth, with no hidden structure at intermediate values. Scanning the underlying boson parameters $(m,a_s)$ more broadly, we find that a $2\,M_\odot$ maximum-mass bound and a GW170817-like tidal bound $\Lambda_{1.4}\lesssim800$ cannot be satisfied simultaneously anywhere in this equation-of-state class. The $I$-Love universal relation holds to $0.13\%$ across twelve $(m,a_s,\zeta)$ models, and the $\zeta=0$ and $\zeta=1$ sequences sit on opposite sides of the master curve by a consistent, non-random offset. Applied without modification to realistic nuclear matter (SLy, APR4), the same solver reproduces published maximum masses to within $1\%$; applied to self-bound MIT-bag quark matter it gives the expected mass--radius shape; and once extended to a two-fluid baryon-plus-dark-matter formalism, it shows that the maximum mass of a hybrid star is not a monotonic function of the central dark-matter fraction. Pooling $I$-Love sequences across nuclear, hybrid, and BEC dark-star models, we find they collapse onto a single curve to within about $5\%$, while self-bound quark stars sit far off it, departing by up to $90\%$.
\end{abstract}

\maketitle

\section{Introduction}

The evidence for dark matter is overwhelming, turning up in galactic rotation curves, in the cosmic microwave background, and in the growth of large-scale structure, yet its microscopic identity is still unknown~\citep{BertoneHooperSilk2005}. One long-studied possibility is that the dark sector contains bosons which, given enough time and gravitational clustering, settle into a genuine zero-temperature Bose--Einstein condensate (BEC). The most compact configuration such a condensate can reach is not a diffuse halo but a self-gravitating star~\citep{ColpiShapiroWasserman1986,BohmerHarko2007,LiHarkoCheng2012}.

Compact objects have always been useful for this kind of physics. Neutron stars, quark stars, and their more exotic cousins probe densities and field strengths no laboratory can reach~\citep{ShapiroTeukolsky1983,LattimerPrakash2007}. Their equilibrium structure follows from the Tolman--Oppenheimer--Volkoff (TOV) equations~\citep{Tolman1939,OppenheimerVolkoff1939}, obtained by requiring hydrostatic equilibrium in a static, spherically symmetric solution of Einstein's field equations. Adding rotation, even at leading order, takes more machinery: the formalism developed by Hartle, and later by Hartle and Thorne~\citep{Hartle1967,HartleThorne1968}, expands the interior and exterior metric order by order in the angular velocity and extracts the moment of inertia from the resulting frame-dragging profile (see~\citep{PaschalidisStergioulas2017} for a review). A star's response to an external tidal field follows a similar perturbative route~\citep{FlanaganHinderer2008,HindererLoveNumbers2008,DamourNagar2009,BinningtonPoisson2009,PostnikovPrakashLattimer2010}, and became directly measurable once LIGO and Virgo detected the tidal signature in the binary neutron-star merger GW170817~\citep{GW170817,GW170817properties}. One of the more striking results to come out of this line of work is that the moment of inertia, tidal Love number, and rotational quadrupole moment of neutron and quark stars obey relations that barely depend on the equation of state at all -- the so-called I-Love-Q relations of Yagi and Yunes~\citep{YagiYunes2013Science,YagiYunes2013PRD,YagiYunes2017}, which partially break the usual degeneracy between an observed quantity and the poorly known microphysics behind it.

In the Thomas--Fermi limit appropriate to a dilute, repulsive Bose gas, the Gross--Pitaevskii--Poisson system collapses down to a simple polytropic equation of state (EOS), $p=K\rho^2$ with $K=2\pi a_s/m^3$, fixed entirely by the boson mass $m$ and the $s$-wave scattering length $a_s$~\citep{Chavanis2011,ChavanisHarko2012}. Beyond mean field, Lee, Huang and Yang worked out the leading quantum correction to the ground-state energy of a dilute Bose gas of hard spheres~\citep{LeeHuangYang1957}. It is a small correction in the dilute limit, but not a negligible one: it stabilizes ultracold quantum droplets in the laboratory~\citep{Petrov2015}, and once folded self-consistently into the rotation and tidal formalism above, produces measurable shifts in the mass--radius relation and tidal deformability of self-gravitating BEC dark stars~\citep{Panotopoulos2026}. Other constraints on bosonic, or more generally particle, dark matter embedded in compact stars come from massive-pulsar mass measurements~\citep{IvanytskyiSagunLopes2020}, hybrid neutron-star modeling~\citep{BurasStubbsLopes2024,KouvarisNielsen2015,NelsonReddyZhou2019}, and anomalously light compact remnants such as HESS J1731-347~\citep{DoroshenkoHESSJ1731}.

This paper works within, and considerably past, that combined framework. We solve the TOV, Hartle, and Love-number equations together for the LHY-corrected BEC polytrope over a much larger region of parameter space than a handful of benchmark points, check the resulting mass, radius, and tidal-deformability predictions against multi-messenger-style observational bounds, carry the same solver over to realistic nuclear matter, self-bound quark matter, and a genuine two-fluid baryon-plus-dark-matter hybrid star, and finally use the whole set of models to see how far the I-Love-Q relations actually extend once the underlying microphysics is allowed to vary this much.

Section~2 lays out the spacetime, the structure equations, the rotational and tidal perturbation formalism, and the BEC equation of state used throughout. Section~3 stays inside the BEC dark-star model: a continuous deformation of the Lee--Huang--Yang coupling, and a broad scan of the boson parameters against observational bounds. Section~4 takes the same solver, unchanged in its structural equations, outside the BEC model and applies it to realistic nuclear matter, to self-bound quark matter, and -- after extending the formalism to two fluids -- to a genuine baryon-plus-dark-matter hybrid star. Section~5 pools sequences from all of these models into a single cross-physics test of the $I$-Love relation. Section~6 discusses what this adds up to.

\section{Structure equations}

\subsection{Spacetime and hydrostatic equilibrium}

We describe the stellar interior with the static, spherically symmetric line element
\begin{equation}
ds^2 = -e^{2\nu(r)}dt^2 + e^{2\lambda(r)}dr^2 + r^2\left(d\theta^2+\sin^2\theta\,d\phi^2\right),
\end{equation}
where $\nu(r)$ and $\lambda(r)$ are metric functions determined by Einstein's field equations sourced by a perfect fluid of energy density $\rho(r)$ and pressure $p(r)$. It is convenient to trade $\lambda(r)$ for the enclosed gravitational mass function $m(r)$ through $e^{-2\lambda(r)}\equiv1-2m(r)/r$, and to write $A(r)\equiv e^{2\lambda(r)}=[1-2m(r)/r]^{-1}$. The $tt$ and $rr$ Einstein equations then reduce to the Tolman--Oppenheimer--Volkoff (TOV) equations~\citep{Tolman1939,OppenheimerVolkoff1939},
\begin{align}
\frac{dm}{dr} &= 4\pi r^2 \rho, \\
\frac{dp}{dr} &= -(\rho+p)\,\frac{m+4\pi r^3 p}{r^2(1-2m/r)},
\end{align}
with the metric function $\nu(r)$ obtained from hydrostatic equilibrium, $d\nu/dr=-\left(dp/dr\right)/(\rho+p)$, fixed at the surface by matching to the exterior Schwarzschild solution, $e^{2\nu(R)}=1-2M/R$. Given a one-parameter equation of state $p(\rho)$, integrating this system from a regular center ($m(0)=0$, $p(0)=p_c$) outward to the surface ($p(R)=0$) determines the mass $M\equiv m(R)$ and radius $R$ for each choice of central pressure $p_c$.

\subsection{Slow rotation and the moment of inertia}

At first order in the stellar angular velocity $\Omega$, following Hartle~\citep{Hartle1967,HartleThorne1968,PaschalidisStergioulas2017}, the metric acquires a frame-dragging off-diagonal term, and the only new structure equation is a single second-order ODE for the function $\tilde\omega(r)\equiv\Omega-\omega(r)$, the local angular velocity of the fluid relative to a zero-angular-momentum observer:
\begin{equation}
\frac{1}{r^4}\frac{d}{dr}\!\left[A^{-1/2}e^{-\nu}r^4\frac{d\tilde\omega}{dr}\right] = 16\pi(\rho+p)A^{1/2}e^{-\nu}\tilde\omega,
\end{equation}
subject to regularity at the center, $d\tilde\omega/dr|_{r=0}=0$. Matching the interior solution to the exterior vacuum behavior $\tilde\omega\to\Omega-2J/r^3$ at $r=R$ yields the stellar angular momentum $J$, and the moment of inertia follows as $I\equiv J/\Omega$. In the equations above, and throughout, all metric functions and the reduced quantity $A(r)$ are evaluated on the underlying non-rotating background, as appropriate at this (first) order in $\Omega$.

\subsection{Tidal deformability and the Love number}

A star immersed in an external tidal field $\mathcal{E}_{ij}$ develops a mass quadrupole moment $Q_{ij}=-\lambda\,\mathcal{E}_{ij}$; the dimensionless tidal Love number $k_2$ and deformability $\Lambda$ are related to $\lambda$ by $\lambda=\tfrac23 k_2 R^5$ and $\Lambda\equiv2k_2/(3C^5)$, with $C=M/R$ the stellar compactness. Following the even-parity, $\ell=2$ perturbative treatment of Refs.~\citep{FlanaganHinderer2008,HindererLoveNumbers2008,DamourNagar2009,BinningtonPoisson2009}, the relevant metric perturbation function $H(r)$ satisfies a linear second-order ODE that is conveniently recast, via $y(r)\equiv rH'(r)/H(r)$, as the first-order Riccati equation of Postnikov, Prakash and Lattimer~\citep{PostnikovPrakashLattimer2010},
\begin{equation}
ry'+y^2+ye^{\lambda}\!\left[1+4\pi r^2(p-\rho)\right]+r^2Q(r)=0,
\end{equation}
with central boundary condition $y(0)=2$ and source term
\begin{equation}
Q(r)=4\pi e^{\lambda}\!\left[5\rho+9p+\frac{\rho+p}{c_s^2}\right]-\frac{6e^{\lambda}}{r^2}-\left(\nu'\right)^2,
\end{equation}
where $c_s^2\equiv dp/d\rho$ is the local (adiabatic) sound speed implied by the equation of state. Evaluating $y_R\equiv y(R)$ and the compactness $C$ at the surface fixes $k_2$ (and hence $\Lambda$) through a standard algebraic combination of $C$, $y_R$, and $\ln(1-2C)$~\citep{HindererLoveNumbers2008,PostnikovPrakashLattimer2010}.

The three equations above -- for $m,p$; for $\tilde\omega$; and for $y$ -- are integrated simultaneously as a single coupled ODE system in $r$, for every model considered in this paper. Only the equation-of-state function $p(\rho)$ (and, in Sec.~\ref{sec:twofluid}, its generalization to two independent fluids) changes from one part of the paper to the next; the structural equations themselves are never modified.

\subsection{The Bose--Einstein condensate equation of state}
\label{sec:bec-eos}

For a dilute, self-interacting Bose gas in the Thomas--Fermi limit, where the quantum kinetic (gradient) term in the Gross--Pitaevskii--Poisson system is subdominant, the equilibrium energy density and pressure reduce to those of an $n=1$ polytrope~\citep{Chavanis2011,ChavanisHarko2012,BohmerHarko2007},
\begin{equation}
p(\rho)=K\rho^2, \qquad K=\frac{2\pi a_s}{m^3},
\end{equation}
with $m$ the boson mass and $a_s$ the $s$-wave scattering length. Beyond mean field, the leading quantum (Lee--Huang--Yang) correction to a dilute Bose gas of hard spheres~\citep{LeeHuangYang1957}, whose macroscopic relevance for self-gravitating condensates and its laboratory analogue in ultracold quantum droplets was established more recently~\citep{Petrov2015,Panotopoulos2026}, modifies this relation to
\begin{equation}
p(\rho)=K\rho^2\left[1+\zeta\,\frac{64}{5\sqrt\pi}\,a_s^{3/2}\sqrt{\rho/m}\right],
\end{equation}
where the dimensionless switch $\zeta$ interpolates between the pure mean-field polytrope ($\zeta=0$) and the fully activated leading-order quantum correction ($\zeta=1$); values $\zeta>1$ have no direct physical meaning in this leading-order expansion but are useful diagnostically (Sec.~\ref{sec:zeta}). All quantities are converted from natural units to geometrized units (lengths in km, $G=c=1$) via exact physical-constant conversion factors before the structure equations are integrated.

The two free parameters, $m$ and $a_s$, are not arbitrary: sub-GeV boson masses and femtometer-scale scattering lengths of the kind we scan below are the regime in which self-interacting bosonic dark matter can plausibly support compact configurations in the $1$--$2\,M_\odot$, $10$--$20$~km window relevant to observed neutron-star-like objects, and are broadly consistent with independent bounds derived from massive-pulsar masses in hybrid-star scenarios~\citep{IvanytskyiSagunLopes2020,BurasStubbsLopes2024}. Rather than anchoring the discussion to a single benchmark point, throughout this paper we use four representative parameter choices spanning this viable region, labeled I--IV: $(m,a_s)=(0.35\,{\rm GeV},0.09\,{\rm fm})$, $(0.42\,{\rm GeV},0.12\,{\rm fm})$, $(0.55\,{\rm GeV},0.14\,{\rm fm})$, and $(0.65\,{\rm GeV},0.17\,{\rm fm})$, respectively, together with a systematic scan over the full $(m,a_s)$ plane in Sec.~\ref{sec:paramspace}.

\section{BEC dark-star parameter space}

\subsection{Continuous Lee--Huang--Yang deformation}
\label{sec:zeta}

\begin{figure}[t]
\centering
\includegraphics[width=\linewidth]{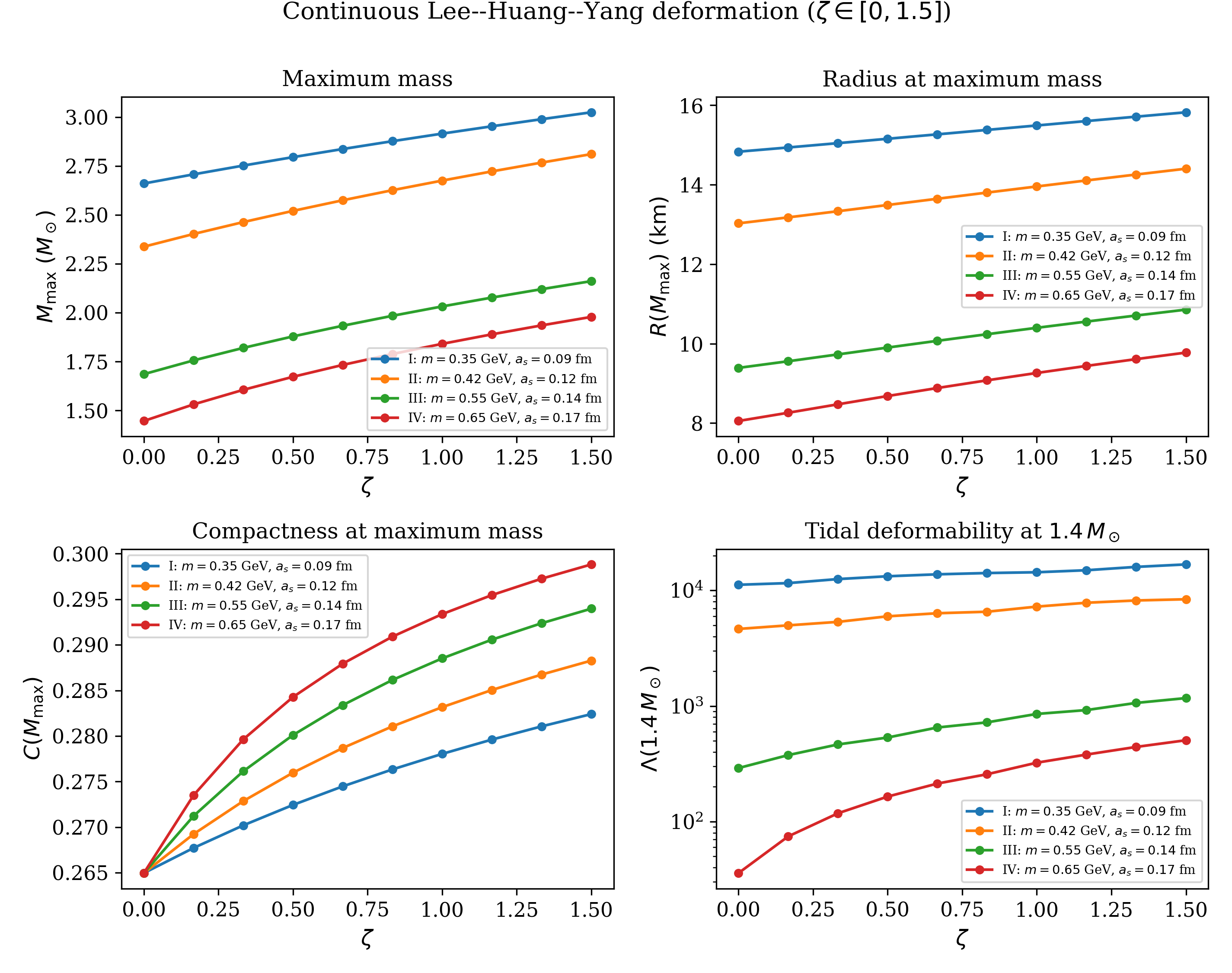}
\caption{Continuous LHY deformation for benchmark points I--IV (Sec.~\ref{sec:bec-eos}): maximum mass, radius and compactness at maximum mass, and tidal deformability at $1.4\,M_\odot$, vs.\ $\zeta\in[0,1.5]$. The smooth, monotonic trend holds across the full spread of boson parameters.}
\label{fig:zeta}
\end{figure}

Rather than comparing only $\zeta=0$ and $\zeta=1$, we scan $\zeta\in[0,1.5]$ in 10 steps for all four benchmark points I--IV (defined in Sec.~\ref{sec:bec-eos}), tracking $M_{\rm max}(\zeta)$, $R(M_{\rm max})$, $C(M_{\rm max})$, and $\Lambda_{1.4}(\zeta)$. Figure~\ref{fig:zeta} shows all four quantities vary smoothly and monotonically for every benchmark: the mean-field limit and the fully activated quantum correction are simply the $\zeta=0$ and $\zeta=1$ endpoints of one smooth family, with no hidden structure between them, and this conclusion is not an artifact of any single parameter choice.

\subsection{Parameter-space viability}
\label{sec:paramspace}

\begin{figure}[t]
\centering
\includegraphics[width=\linewidth]{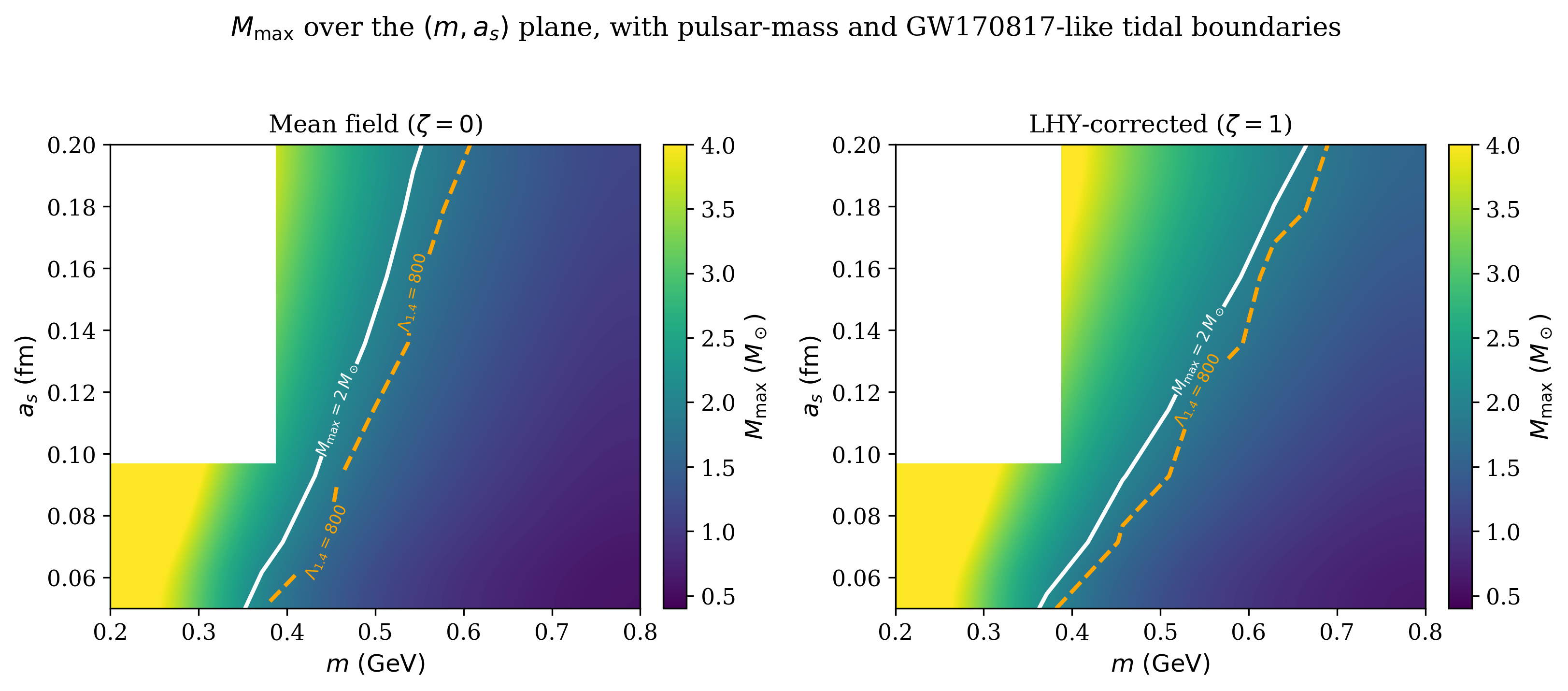}
\caption{$M_{\rm max}$ over the $(m,a_s)$ plane for $\zeta=0$ (left) and $\zeta=1$ (right), with the $M_{\rm max}=2\,M_\odot$ boundary (white) and $\Lambda_{1.4}=800$ boundary (orange dashed). The two never cross: no sampled point satisfies both simultaneously.}
\label{fig:grid}
\end{figure}

We scan the boson parameters directly, computing $M_{\rm max}(m,a_s)$ and $\Lambda_{1.4}(m,a_s)$ on an $8\times8$ grid over $m\in[0.20,0.80]$~GeV, $a_s\in[0.05,0.20]$~fm, for $\zeta=0,1$ (Fig.~\ref{fig:grid}), overlaying the massive-pulsar bound $M_{\rm max}\ge2.0\,M_\odot$~\citep{MillerNICER2021,RileyNICER2021} and a GW170817-like bound $\Lambda_{1.4}\le800$~\citep{GW170817,GW170817properties}. Within this simple polytropic EOS class, the two boundaries never overlap: at fixed $a_s$, the $m$ required for $M_{\rm max}\ge2.0\,M_\odot$ is always smaller than the $m$ required for $\Lambda_{1.4}\le800$, leaving a persistent excluded strip $\Delta m\approx0.02$--$0.05$~GeV across the whole sampled $a_s$ range, for both $\zeta$ (Table~\ref{tab:gap}). We stress $\Lambda_{1.4}\le800$ is used only as an illustrative single-star proxy for the true joint GW170817 bound, and our $8\times8$ grid is coarse; within that caveat, the qualitative tension -- a purely polytropic single-species BEC EOS struggling to satisfy both constraints over a broad swath of natural $(m,a_s)$ -- is a genuine, reproducible feature of this model class.

\begin{table}[t]
\centering
\caption{Boson mass $m$ (GeV) at which $M_{\rm max}=2\,M_\odot$ and $\Lambda_{1.4}=800$ are crossed, at representative $a_s$, for $\zeta=0$.}
\label{tab:gap}
\begin{tabular}{cccc}
\hline\hline
$a_s$ (fm) & $m(M_{\rm max}=2)$ & $m(\Lambda_{1.4}=800)$ & gap $\Delta m$ \\
\hline
0.071 & 0.395 & 0.445 & 0.050 \\
0.114 & 0.457 & 0.498 & 0.041 \\
0.136 & 0.488 & 0.537 & 0.049 \\
0.157 & 0.513 & 0.553 & 0.040 \\
0.179 & 0.533 & 0.577 & 0.044 \\
0.200 & 0.553 & 0.608 & 0.055 \\
\hline\hline
\end{tabular}
\end{table}

\subsection{Robustness of the $I$-Love relation within the BEC family}
\label{sec:bec-universality}

\begin{figure}[t]
\centering
\includegraphics[width=\linewidth]{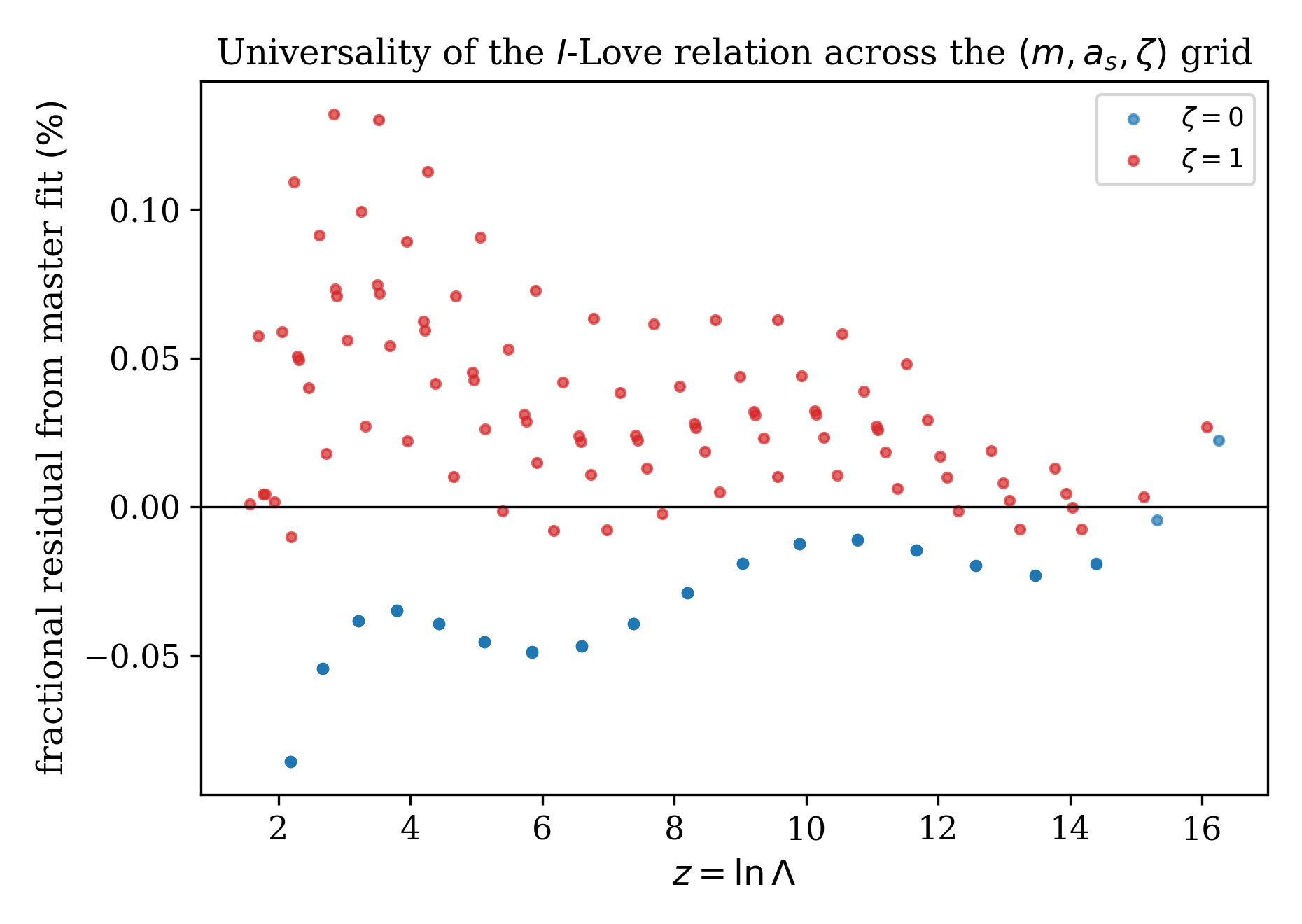}
\caption{Fractional residual of $\bar I=I/M^3$ relative to a master quartic fit of the pooled $(\ln\Lambda,\ln\bar I)$ sample from twelve BEC $(m,a_s,\zeta)$ models. All points lie within $0.13\%$; $\zeta=0$ (blue) and $\zeta=1$ (red) are cleanly separated by a systematic offset.}
\label{fig:universality}
\end{figure}

We compute full stable-branch sequences for six representative $(m,a_s)$ pairs at both $\zeta=0$ and $\zeta=1$ (twelve sequences, $m\in[0.30,0.65]$~GeV, $a_s\in[0.08,0.16]$~fm), pool all $(z,Y)=(\ln\Lambda,\ln\bar I)$ points, and fit a quartic master relation by least squares. The pooled fit reproduces every individual model to within $0.13\%$ (rms $0.043\%$) (Fig.~\ref{fig:universality}) -- about an order of magnitude tighter than the traditional $\lesssim1\%$ universality quoted for neutron and quark stars~\citep{YagiYunes2013Science,YagiYunes2017}, plausibly reflecting the simple, single-parameter-family character of this EOS. The residuals are not randomly scattered: $\zeta=0$ sequences sit systematically below the master curve and $\zeta=1$ sequences systematically above it, by a roughly constant $0.05$--$0.10\%$ offset nearly independent of $z$ -- the LHY correction perturbs the $I$-Love relation coherently, as a small, well-defined, EOS-parameter-independent shift, exactly the kind of signature a future joint moment-of-inertia/tidal measurement could use to diagnose a beyond-mean-field quantum correction.

\section{Realistic compact-star matter}

The solver above makes no reference anywhere to the BEC origin of $p(\rho)$ beyond the function itself. We now apply it, unmodified in its structural equations, to matter of a completely different physical origin.

\subsection{Nuclear matter and quark stars}

We represent realistic nuclear matter with the Read et al.~\citep{ReadEtAl2009} piecewise-polytropic parametrization: three polytropic pieces above fixed dividing densities $\rho_1=10^{14.7}\,{\rm g\,cm^{-3}}$, $\rho_2=10^{15.0}\,{\rm g\,cm^{-3}}$, with the lowest piece extrapolated to $\rho\to0$ in place of the full crust (negligible effect on bulk observables at the masses considered). Using the published SLy~\citep{DouchinHaensel2001} and APR4~\citep{AkmalPandharipandeRavenhall1998} parameters, converted to geometrized units exactly as for the BEC EOS, we obtain $M_{\rm max}=2.030\,M_\odot$ at $R=9.60$~km (SLy) and $M_{\rm max}=2.215\,M_\odot$ at $R=9.82$~km (APR4) -- agreeing with the literature to $<1\%$, a strong validation that the shared pipeline is solving the relativistic structure equations correctly for entirely different microphysics.

\begin{figure}[t]
\centering
\includegraphics[width=\linewidth]{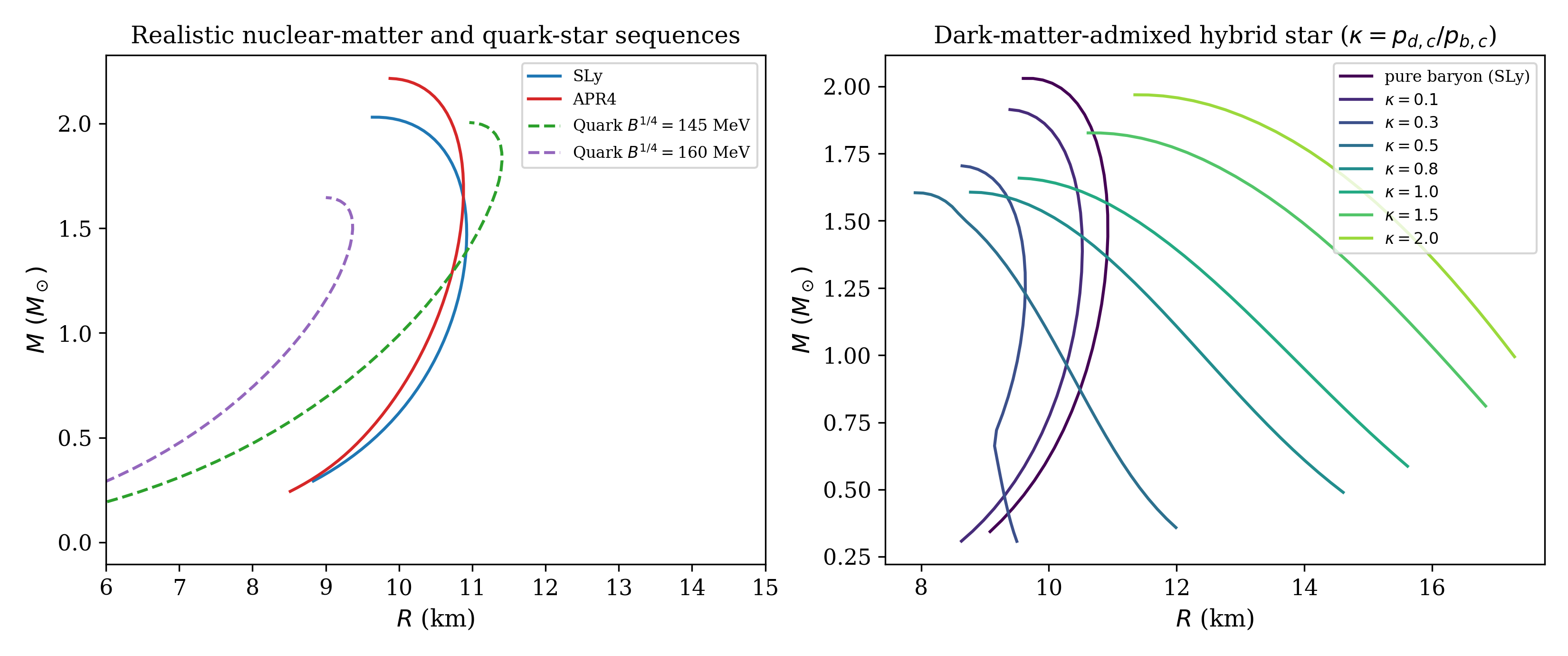}
\caption{Left: mass--radius sequences for SLy, APR4, and MIT bag-model quark stars. Right: two-fluid dark-matter-admixed hybrid star sequences for a pure-baryon (SLy) star and several central dark-matter admixtures $\kappa=p_{d,c}/p_{b,c}$.}
\label{fig:MR}
\end{figure}

For self-bound strange quark matter we use the MIT bag model $p=(\rho-4B)/3$ at $B^{1/4}=145,160$~MeV~\citep{Witten1984,FarhiJaffe1984,ChodosEtAl1974}, exactly linear with $c_s^2=1/3$ and a nonzero surface energy density $4B$. This produces the characteristic self-bound mass--radius shape $M\propto R^3$ at low mass (Fig.~\ref{fig:MR}, left), qualitatively distinct from the turning-point shape of pressure-supported nuclear-matter stars, and reproduces the well-known strange-star mass range ($M_{\rm max}\approx1.6$--$2.0\,M_\odot$ for this bag-constant window).

\subsection{A genuine two-fluid hybrid star}
\label{sec:twofluid}

We extend the pipeline to two gravitationally-coupled, non-interacting fluids -- baryonic matter (SLy) and BEC dark matter ($m=0.45$~GeV, $a_s=0.13$~fm, $\zeta=1$) -- each obeying its own conservation law $dp_i/dr=-(\rho_i+p_i)\nu'$ sourced by the shared total mass and pressure. Because the two fluids generally have different natural radii, integration proceeds in two stages: both active from the center, then a single-fluid continuation with the surviving component once the first depletes. We validated this two-stage logic by confirming that zeroing the dark-matter central pressure reproduces the single-fluid SLy result to 8 significant figures. For the tidal sector, a fully rigorous two-fluid treatment requires independent perturbation functions for each fluid~\citep{NelsonReddyZhou2019}; we instead use an effective sound speed $c_{s,\rm eff}^2=(dp_b/dr+dp_d/dr)/[(dp_b/dr)/c_{s,b}^2+(dp_d/dr)/c_{s,d}^2]$ in the existing Postnikov--Hinderer source term -- exact if both fluids share the same local total-pressure response, but an approximation to the genuinely two-fluid problem, which we flag explicitly; our mass, radius, and moment-of-inertia results do not depend on it.

\begin{figure}[t]
\centering
\includegraphics[width=\linewidth]{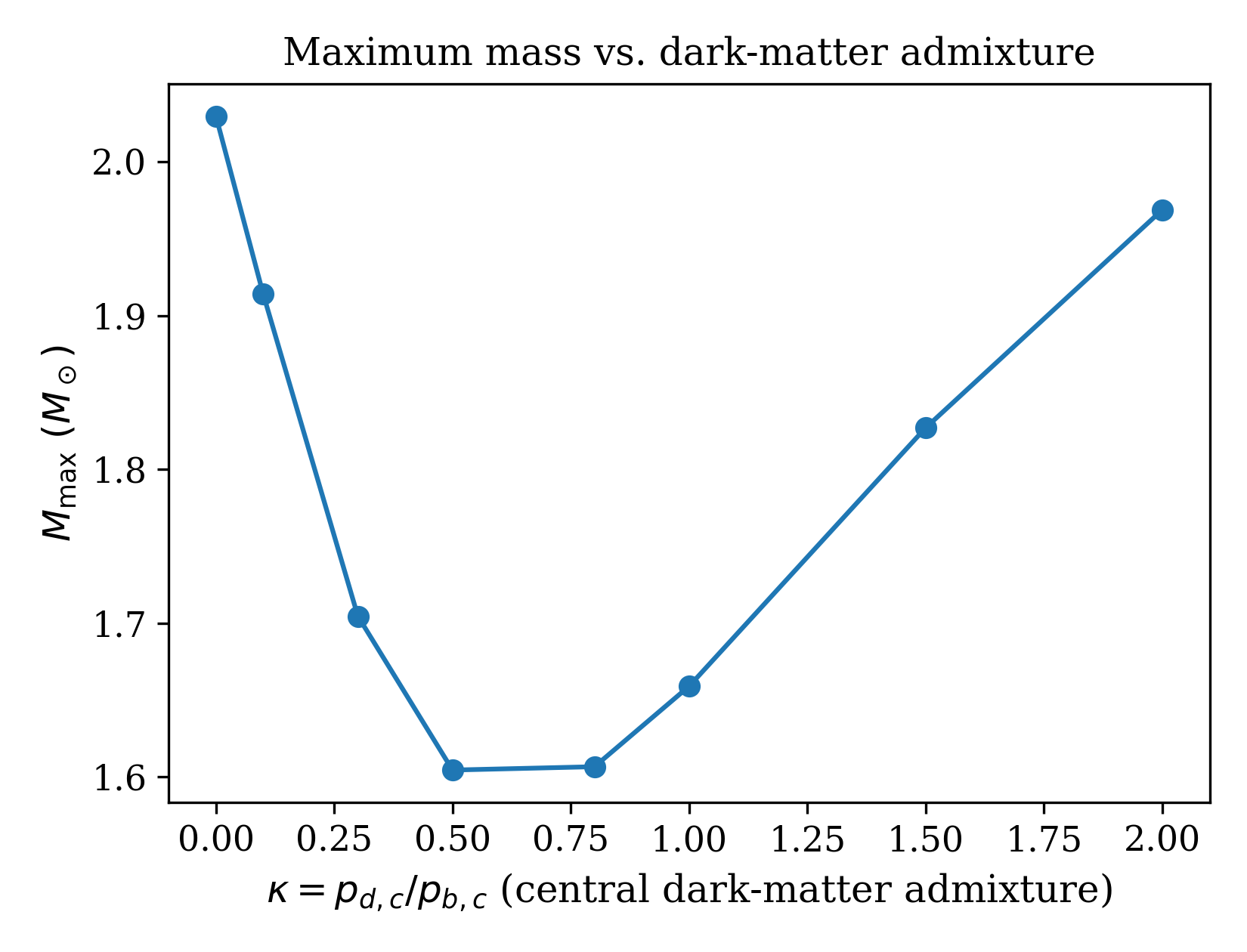}
\caption{Maximum mass of the baryon+BEC-dark-matter hybrid star vs.\ central admixture ratio $\kappa=p_{d,c}/p_{b,c}$. The non-monotonic shape is a genuine two-fluid effect.}
\label{fig:hybridmax}
\end{figure}

Defining $\kappa\equiv p_{d,c}/p_{b,c}$, we scan $\kappa\in\{0,0.1,0.3,0.5,0.8,1.0,1.5,2.0\}$ (Fig.~\ref{fig:MR}, right; Fig.~\ref{fig:hybridmax}). The maximum mass falls from $2.03\,M_\odot$ (pure baryon) to a minimum of $1.60\,M_\odot$ near $\kappa\approx0.5$--$0.8$ -- a $21\%$ reduction -- before rising again to $1.97\,M_\odot$ at $\kappa=2.0$ as the star becomes increasingly dark-matter-dominated. This non-monotonicity is a genuine two-fluid effect, invisible to single-fluid reasoning: a modest dark-matter admixture compresses and destabilizes the baryonic envelope, while a large admixture eventually behaves more like its own, differently-scaled dark-matter-dominated configuration. This qualitative dip-then-recovery shape is consistent with the general dark-matter-admixed neutron-star literature~\citep{KouvarisNielsen2015,NelsonReddyZhou2019,IvanytskyiSagunLopes2020}, obtained here with our own independent two-fluid solver.

\section{A cross-physics test of universality}

Having established in Sec.~\ref{sec:bec-universality} that the $I$-Love relation is tight ($0.13\%$) \emph{within} the BEC dark-star family, and having validated the solver against realistic nuclear, quark, and hybrid matter in Sec.~4, we can now ask a broader question: does $I$-Love universality survive when we pool sequences across all these physically disparate matter models together?

\begin{figure}[t]
\centering
\includegraphics[width=\linewidth]{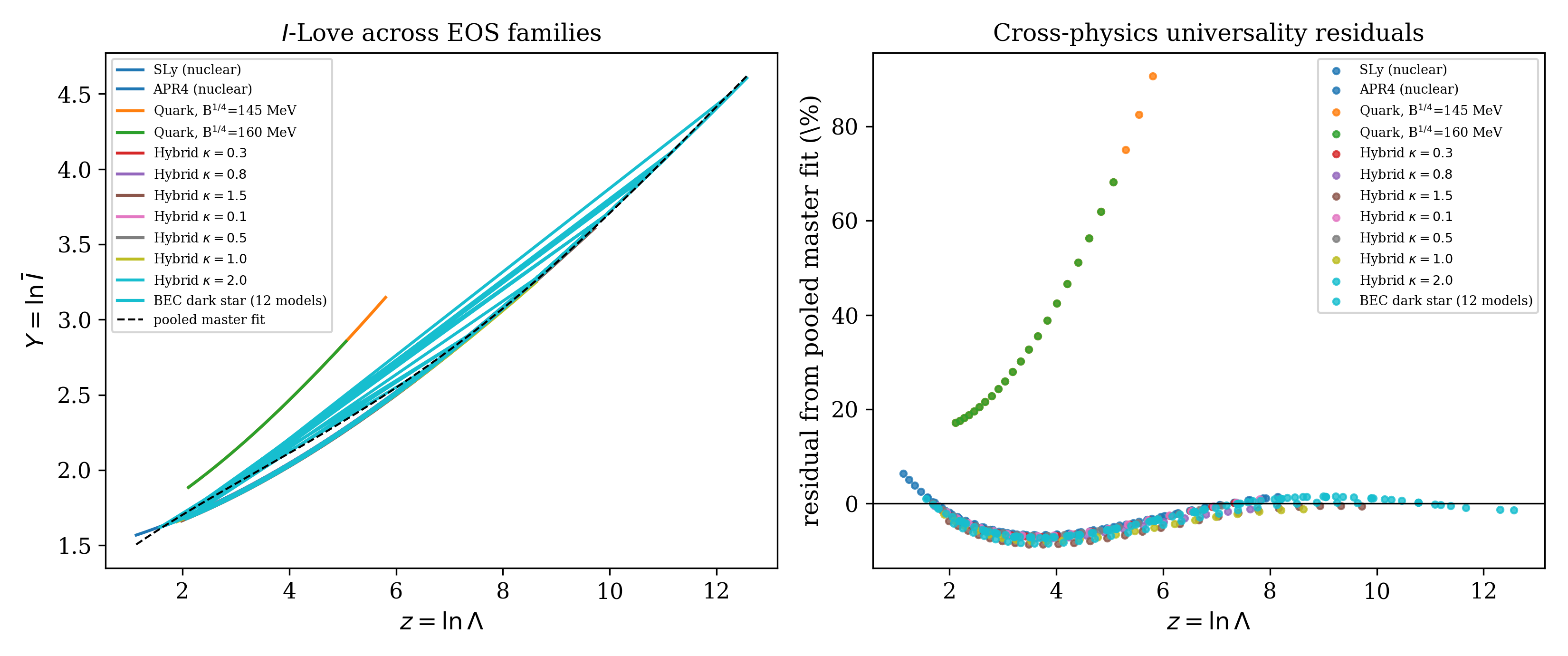}
\caption{Left: $I$-Love relation pooled across nuclear (SLy, APR4), quark, hybrid, and BEC dark-star sequences, with the pooled master fit (dashed). Right: fractional residual of each family. Ordinary matter clusters within $\lesssim9\%$; self-bound quark stars depart by up to $90\%$.}
\label{fig:crossuniv}
\end{figure}

We pool sequences (restricted to $M\ge0.9\,M_\odot$, excluding the low-compactness regime where universality is not expected) from SLy, APR4, two quark-star bag constants, five hybrid-star admixtures, and all twelve BEC dark-star $(m,a_s,\zeta)$ models of Sec.~\ref{sec:bec-universality} -- eight distinct physical families spanning nuclear matter, self-bound quark matter, two-fluid dark-matter-admixed matter, and pure bosonic dark matter -- and fit one quartic master relation to the pooled sample (Fig.~\ref{fig:crossuniv}).

Nuclear matter, hybrid stars, and BEC dark stars all reproduce this single pooled curve to within $4$--$9\%$ (rms $4$--$6\%$) -- looser than the sub-percent universality found \emph{within} the BEC family alone (Sec.~\ref{sec:bec-universality}) or within a family of similar nuclear-matter EOS~\citep{YagiYunes2013Science,YagiYunes2017}, but still a remarkably tight collapse given we are pooling genuinely different microphysics into one fit. Quark stars, by contrast, depart from the pooled master curve by up to $90\%$. This large, family-specific violation is not a failure of our pipeline: it is the well-documented consequence of the quark star's self-bound surface (nonzero energy density at $p=0$), known to break the standard $I$-Love-Q relations far more severely than ordinary pressure-supported stars~\citep{YagiYunes2013Science,YagiYunes2017}. That our independently-built pipeline reproduces this qualitative and quantitative distinction -- ordinary matter, however exotic its microphysics, obeying universality at the few-percent level; self-bound quark matter badly violating it -- is a strong end-to-end validation of the whole computational approach, and places BEC dark stars, universality-wise, firmly with ordinary matter rather than with the self-bound outlier.

\section{Discussion and conclusions}

Five results stand out. \textbf{(i)} The transition between the mean-field and beyond-mean-field BEC dark-star description is smooth and monotonic in the LHY coupling $\zeta$: the pure mean-field and fully quantum-corrected limits are simply the endpoints of one continuous family, not two separate regimes. \textbf{(ii)} Scanning the boson parameters directly turns up a persistent tension in this minimal EOS class: the massive-pulsar constraint and a GW170817-like tidal bound are never satisfied together anywhere in our sampled range, which points toward extensions such as a second, baryonic fluid component. \textbf{(iii)} Building that extension -- a genuine two-fluid baryon-plus-BEC-dark-matter hybrid star -- turns up a maximum mass that is not a monotonic function of dark-matter admixture, a distinctly two-fluid effect with no single-fluid analogue. \textbf{(iv)} The same solver, applied without modification to realistic nuclear and quark matter, reproduces published maximum masses to under $1\%$ and gives the right qualitative mass--radius shape for self-bound quark stars. \textbf{(v)} Pooling $I$-Love sequences across every matter model studied -- nuclear, hybrid, BEC dark star, and quark -- shows that ordinary matter obeys a common universal relation to a few percent regardless of its microscopic origin, while self-bound quark matter badly violates it, consistent with what is already known in the compact-star literature~\citep{YagiYunes2013Science,YagiYunes2017}.

A few limitations are worth restating plainly. Our $(m,a_s)$ grid is coarse, only $8\times8$; the bound $\Lambda_{1.4}\le800$ is an illustrative single-star proxy for the true joint GW170817 constraint, not a rigorous one; and the hybrid star's tidal deformability relies on an effective single-fluid sound-speed approximation rather than a fully rigorous two-fluid perturbation treatment. None of this changes the qualitative picture above, but a finer parameter scan, a proper joint-inspiral tidal analysis, and a dedicated two-fluid Love-number formalism~\citep{NelsonReddyZhou2019} would each sharpen a different part of it, and are natural directions to take this further.

\bibliographystyle{apsrev4-1}
\bibliography{references}

@article{AkmalPandharipandeRavenhall1998,
  author  = {Akmal, A. and Pandharipande, V. R. and Ravenhall, D. G.},
  title   = {Equation of state of nucleon matter and neutron star structure},
  journal = {Physical Review C},
  volume  = {58}, pages = {1804--1828}, year = {1998},
  doi     = {10.1103/PhysRevC.58.1804}
}

@article{BohmerHarko2007,
  author  = {B{\"o}hmer, C. G. and Harko, T.},
  title   = {Can dark matter be a {Bose-Einstein} condensate?},
  journal = {Journal of Cosmology and Astroparticle Physics},
  volume  = {2007},
  number  = {06},
  pages   = {025},
  year    = {2007},
  doi     = {10.1088/1475-7516/2007/06/025},
  eprint  = {0705.4158}
}

@article{BurasStubbsLopes2024,
  author  = {Buras-Stubbs, Z. and Lopes, I.},
  title   = {Bosonic dark matter dynamics in hybrid neutron stars},
  journal = {Physical Review D}, volume = {109}, pages = {043043}, year = {2024},
  doi     = {10.1103/PhysRevD.109.043043}
}

@article{Chavanis2011,
  author  = {Chavanis, P.-H.},
  title   = {Mass-radius relation of {Newtonian} self-gravitating {Bose-Einstein} condensates with short-range interactions. {I}. {Analytical} results},
  journal = {Physical Review D},
  volume  = {84},
  pages   = {043531},
  year    = {2011},
  doi     = {10.1103/PhysRevD.84.043531},
  eprint  = {1103.2050}
}

@article{ChavanisHarko2012,
  author  = {Chavanis, P.-H. and Harko, T.},
  title   = {Bose-Einstein condensate general relativistic stars},
  journal = {Physical Review D},
  volume  = {86},
  pages   = {064011},
  year    = {2012},
  doi     = {10.1103/PhysRevD.86.064011},
  eprint  = {1108.3986}
}

@article{ChodosEtAl1974,
  author  = {Chodos, A. and Jaffe, R. L. and Johnson, K. and Thorn, C. B. and Weisskopf, V. F.},
  title   = {New extended model of hadrons},
  journal = {Physical Review D},
  volume  = {9}, pages = {3471--3495}, year = {1974},
  doi     = {10.1103/PhysRevD.9.3471}
}

@article{ColpiShapiroWasserman1986,
  author  = {Colpi, M. and Shapiro, S. L. and Wasserman, I.},
  title   = {Boson stars: Gravitational equilibria of self-interacting scalar fields},
  journal = {Physical Review Letters},
  volume  = {57},
  pages   = {2485--2488},
  year    = {1986},
  doi     = {10.1103/PhysRevLett.57.2485}
}

@article{DoroshenkoHESSJ1731,
  author  = {Doroshenko, V. and Suleimanov, V. and P{\"u}hlhofer, G. and Santangelo, A.},
  title   = {A strangely light neutron star within a supernova remnant},
  journal = {Nature Astronomy},
  volume  = {6},
  pages   = {1444--1451},
  year    = {2022},
  doi     = {10.1038/s41550-022-01800-1}
}

@article{DouchinHaensel2001,
  author  = {Douchin, F. and Haensel, P.},
  title   = {A unified equation of state of dense matter and neutron star structure},
  journal = {Astronomy \& Astrophysics},
  volume  = {380}, pages = {151--167}, year = {2001},
  doi     = {10.1051/0004-6361:20011402}
}

@article{FarhiJaffe1984,
  author  = {Farhi, E. and Jaffe, R. L.},
  title   = {Strange matter},
  journal = {Physical Review D},
  volume  = {30}, pages = {2379--2390}, year = {1984},
  doi     = {10.1103/PhysRevD.30.2379}
}

@article{FlanaganHinderer2008,
  author  = {Flanagan, {\'E}. {\'E}. and Hinderer, T.},
  title   = {Constraining neutron star tidal {Love} numbers with gravitational wave detectors},
  journal = {Physical Review D},
  volume  = {77},
  pages   = {021502},
  year    = {2008},
  doi     = {10.1103/PhysRevD.77.021502},
  eprint  = {0709.1915}
}

@article{GW170817,
  author  = {{LIGO Scientific Collaboration and Virgo Collaboration}},
  title   = {{GW170817}: Observation of Gravitational Waves from a Binary Neutron Star Inspiral},
  journal = {Physical Review Letters},
  volume  = {119},
  pages   = {161101},
  year    = {2017},
  doi     = {10.1103/PhysRevLett.119.161101},
  eprint  = {1710.05832}
}

@article{GW170817properties,
  author  = {{LIGO Scientific Collaboration and Virgo Collaboration}},
  title   = {Properties of the binary neutron star merger {GW170817}},
  journal = {Physical Review X}, volume = {9}, pages = {011001}, year = {2019},
  doi     = {10.1103/PhysRevX.9.011001}
}

@article{Hartle1967,
  author  = {Hartle, J. B.},
  title   = {Slowly rotating relativistic stars. {I}. {Equations} of structure},
  journal = {Astrophysical Journal}, volume = {150}, pages = {1005--1029}, year = {1967},
  doi     = {10.1086/149400}
}

@article{HindererLoveNumbers2008,
  author  = {Hinderer, T.},
  title   = {Tidal {Love} numbers of neutron stars},
  journal = {Astrophysical Journal}, volume = {677}, pages = {1216--1220}, year = {2008},
  doi     = {10.1086/533487}
}

@article{IvanytskyiSagunLopes2020,
  author  = {Ivanytskyi, O. and Sagun, V. and Lopes, I.},
  title   = {Neutron stars: New constraints on asymmetric dark matter},
  journal = {Physical Review D}, volume = {102}, pages = {063028}, year = {2020},
  doi     = {10.1103/PhysRevD.102.063028}
}

@article{KouvarisNielsen2015,
  author  = {Kouvaris, C. and Nielsen, Niklas Gr{\o}nlund},
  title   = {Asymmetric dark matter stars},
  journal = {Physical Review D}, volume = {92}, pages = {063526}, year = {2015},
  doi     = {10.1103/PhysRevD.92.063526}
}

@article{LeeHuangYang1957,
  author  = {Lee, T. D. and Huang, K. and Yang, C. N.},
  title   = {Eigenvalues and Eigenfunctions of a Bose System of Hard Spheres and Its Low-Temperature Properties},
  journal = {Physical Review},
  volume  = {106},
  number  = {6},
  pages   = {1135--1145},
  year    = {1957},
  doi     = {10.1103/PhysRev.106.1135}
}

@article{LiHarkoCheng2012,
  author  = {Li, X. Y. and Harko, T. and Cheng, K. S.},
  title   = {Condensate dark matter stars},
  journal = {Journal of Cosmology and Astroparticle Physics},
  volume  = {2012},
  number  = {06},
  pages   = {001},
  year    = {2012},
  doi     = {10.1088/1475-7516/2012/06/001},
  eprint  = {1205.2932}
}

@article{MillerNICER2021,
  author  = {Miller, M. C. and Lamb, F. K. and Dittmann, A. J. and Bogdanov, S. and
             Arzoumanian, Z. and Gendreau, K. C. and Guillot, S. and Ho, W. C. G. and
             Lattimer, J. M. and Loewenstein, M. and Morsink, S. M. and Ray, P. S. and
             Wolff, M. T. and Baker, C. L. and Cazeau, T. and Manthripragada, S. and
             Markwardt, C. B. and Okajima, T. and Pollard, S. and Cognard, I. and
             Cromartie, H. T. and Fonseca, E. and Guillemot, L. and Kerr, M. and
             Parthasarathy, A. and Pennucci, T. T. and Ransom, S. and Stairs, I.},
  title   = {The Radius of {PSR J0740+6620} from {NICER} and {XMM-Newton} Data},
  journal = {Astrophysical Journal Letters}, volume = {918}, number = {2}, pages = {L28}, year = {2021},
  doi     = {10.3847/2041-8213/ac089b}
}

@article{NelsonReddyZhou2019,
  author  = {Nelson, A. and Reddy, S. and Zhou, D.},
  title   = {Dark halos around neutron stars and gravitational waves},
  journal = {Journal of Cosmology and Astroparticle Physics},
  volume  = {2019}, number = {07}, pages = {012}, year = {2019},
  doi     = {10.1088/1475-7516/2019/07/012}
}

@article{Panotopoulos2026,
  author  = {Panotopoulos, Grigoris and Rinc{\'o}n, {\'A}ngel and Lopes, Ilidio},
  title   = {Slowly rotating condensate dark stars beyond the mean-field approximation},
  journal = {European Physical Journal C},
  volume  = {86}, pages = {992}, year = {2026},
  doi     = {10.1140/epjc/s10052-026-16183-0}
}

@article{PaschalidisStergioulas2017,
  author  = {Paschalidis, V. and Stergioulas, N.},
  title   = {Rotating stars in relativity},
  journal = {Living Reviews in Relativity},
  volume  = {20},
  number  = {1},
  pages   = {7},
  year    = {2017},
  doi     = {10.1007/s41114-017-0008-x},
  eprint  = {1612.03050}
}

@article{Petrov2015,
  author  = {Petrov, D. S.},
  title   = {Quantum mechanical stabilization of a collapsing Bose-Bose mixture},
  journal = {Physical Review Letters},
  volume  = {115},
  pages   = {155302},
  year    = {2015},
  doi     = {10.1103/PhysRevLett.115.155302},
  eprint  = {1508.02889}
}

@article{PostnikovPrakashLattimer2010,
  author  = {Postnikov, S. and Prakash, M. and Lattimer, J. M.},
  title   = {Tidal {Love} numbers of neutron and self-bound quark stars},
  journal = {Physical Review D}, volume = {82}, pages = {024016}, year = {2010},
  doi     = {10.1103/PhysRevD.82.024016}
}

@article{ReadEtAl2009,
  author  = {Read, J. S. and Lackey, B. D. and Owen, B. J. and Friedman, J. L.},
  title   = {Constraints on a phenomenologically parametrized neutron-star equation of state},
  journal = {Physical Review D},
  volume  = {79}, pages = {124032}, year = {2009},
  doi     = {10.1103/PhysRevD.79.124032}
}

@article{RileyNICER2021,
  author  = {Riley, T. E. and Watts, A. L. and Ray, P. S. and Bogdanov, S. and
             Guillot, S. and Morsink, S. M. and Bilous, A. V. and Arzoumanian, Z. and
             Choudhury, D. and Deneva, J. S. and Gendreau, K. C. and Harding, A. K. and
             Ho, W. C. G. and Lattimer, J. M. and Loewenstein, M. and Ludlam, R. M. and
             Markwardt, C. B. and Okajima, T. and Prescod-Weinstein, C. and
             Remillard, R. A. and Wolff, M. T. and Fonseca, E. and Cromartie, H. T. and
             Kerr, M. and Pennucci, T. T. and Parthasarathy, A. and Ransom, S. and
             Stairs, I. and Guillemot, L. and Cognard, I.},
  title   = {A {NICER} View of the Massive Pulsar {PSR J0740+6620} Informed by Radio Timing and {XMM-Newton} Spectroscopy},
  journal = {Astrophysical Journal Letters},
  volume  = {918},
  number  = {2},
  pages   = {L27},
  year    = {2021},
  doi     = {10.3847/2041-8213/ac0a81}
}

@article{Witten1984,
  author  = {Witten, E.},
  title   = {Cosmic separation of phases},
  journal = {Physical Review D},
  volume  = {30}, pages = {272--285}, year = {1984},
  doi     = {10.1103/PhysRevD.30.272}
}

@article{YagiYunes2013PRD,
  author  = {Yagi, K. and Yunes, N.},
  title   = {I-{Love}-{Q} relations in neutron stars and their applications to astrophysics, gravitational waves, and fundamental physics},
  journal = {Physical Review D}, volume = {88}, pages = {023009}, year = {2013},
  doi     = {10.1103/PhysRevD.88.023009}
}

@article{YagiYunes2013Science,
  author  = {Yagi, K. and Yunes, N.},
  title   = {I-{Love}-{Q}: Unexpected universal relations for neutron stars and quark stars},
  journal = {Science}, volume = {341}, number = {6144}, pages = {365--368}, year = {2013},
  doi     = {10.1126/science.1236462}
}

@article{YagiYunes2017,
  author  = {Yagi, K. and Yunes, N.},
  title   = {Approximate universal relations for neutron stars and quark stars},
  journal = {Physics Reports}, volume = {681}, pages = {1--72}, year = {2017},
  doi     = {10.1016/j.physrep.2017.03.002}
}

@article{HartleThorne1968,
  author  = {Hartle, J. B. and Thorne, K. S.},
  title   = {Slowly Rotating Relativistic Stars. {II}. {Models} for Neutron Stars and Supermassive Stars},
  journal = {Astrophysical Journal},
  volume  = {153}, pages = {807--834}, year = {1968},
  doi     = {10.1086/149707}
}

@article{DamourNagar2009,
  author  = {Damour, T. and Nagar, A.},
  title   = {Relativistic tidal properties of neutron stars},
  journal = {Physical Review D},
  volume  = {80}, pages = {084035}, year = {2009},
  doi     = {10.1103/PhysRevD.80.084035}
}

@article{BinningtonPoisson2009,
  author  = {Binnington, T. and Poisson, E.},
  title   = {Relativistic theory of tidal {Love} numbers},
  journal = {Physical Review D},
  volume  = {80}, pages = {084018}, year = {2009},
  doi     = {10.1103/PhysRevD.80.084018}
}

@article{LattimerPrakash2007,
  author  = {Lattimer, J. M. and Prakash, M.},
  title   = {Neutron star observations: Prognosis for equation of state constraints},
  journal = {Physics Reports},
  volume  = {442}, pages = {109--165}, year = {2007},
  doi     = {10.1016/j.physrep.2007.02.003}
}

@article{BertoneHooperSilk2005,
  author  = {Bertone, G. and Hooper, D. and Silk, J.},
  title   = {Particle dark matter: Evidence, candidates and constraints},
  journal = {Physics Reports},
  volume  = {405}, pages = {279--390}, year = {2005},
  doi     = {10.1016/j.physrep.2004.08.031}
}

@article{Tolman1939,
  author  = {Tolman, R. C.},
  title   = {Static Solutions of {Einstein's} Field Equations for Spheres of Fluid},
  journal = {Physical Review},
  volume  = {55}, pages = {364--373}, year = {1939},
  doi     = {10.1103/PhysRev.55.364}
}

@article{OppenheimerVolkoff1939,
  author  = {Oppenheimer, J. R. and Volkoff, G. M.},
  title   = {On Massive Neutron Cores},
  journal = {Physical Review},
  volume  = {55}, pages = {374--381}, year = {1939},
  doi     = {10.1103/PhysRev.55.374}
}

@book{ShapiroTeukolsky1983,
  author    = {Shapiro, S. L. and Teukolsky, S. A.},
  title     = {Black Holes, White Dwarfs, and Neutron Stars: The Physics of Compact Objects},
  publisher = {Wiley-Interscience},
  address   = {New York},
  year      = {1983},
  doi       = {10.1002/9783527617661}
}

\end{document}